POWER GENERATION I ENERGY SUPPLY

Future technologies in power generation

# Electricity in international comparison

AXEL KLEIDON | HARALD LESCH | RUSS CONSER

*Which technologies are currently booming in power generation? The answer can clearly be seen in the trends in global electricity generation data. Analysing this data using the theory of diffusion of innovations reveals how photovoltaics and wind power are gaining ground worldwide. Other technologies—especially coal and nuclear power—are being displaced, with Germany playing a pioneering role.*

Is nuclear energy experiencing a revival? Is China investing mainly in coal-fired power plants? And is Germany on a special path with regard to its energy policy? These questions can be answered objectively by looking at the electricity generation data from recent years. We want to do this here with a data set on global electricity generation [1], with a particular focus on the technologies that are currently gaining ground. To do this, we use the theory of diffusion of innovations from the social sciences [2]. It assumes that the development and spread of innovations can be described as logistic growth. The development dynamics described by this theory combine an initial exponential growth in the spread of innovation with the capacity that limits the spread of innovation. This is a development trend that can be found in many processes.

With the help of this theory, we want to examine electricity generation in Germany and compare it with developments in other countries and regions worldwide. The aim is to identify which innovations are currently gaining ground in electricity generation. We will then address the question of why certain technologies are gaining ground right now by linking these developments to interactions and feedback loops. Finally, we will return to the questions posed at the beginning and recognise that Germany is not playing a special role, but rather a pioneering role. We will link this to developments in electricity prices in Germany.

### How innovations spread in the market

The spread of innovation in biological, ecological, and social systems has long been described in the form of an "S-curve," which represents logistic growth (Figure 1a). Logistic growth is characterised by two parameters: a relative growth rate $r$ – which leads to initial exponential growth – and the aforementioned capacity $K$, which limits growth to a maximum. Examples of such logistic growth include the spread of epidemics, new products such as smartphones, or the spread of technologies for electricity generation. In these examples, the capacity represents, for example, the population that can be infected, the market that can be saturated, or a country's total power generation. This description originally comes from population biology, where $r$ describes the growth rate of a population and $K$ describes the capacity of the environment to sustain that population.

Mathematically, logistic growth is represented by a differential equation that describes the spread of an innovation $N$ as follows:

$$\frac{dN}{dt} = r \cdot N \cdot \left(1 - \frac{N}{K}\right) \qquad (1)$$

Here, $dN/dt$ describes the growth (or decline) of an innovation over time, $r$ is the relative growth rate, and $K$ is the capacity. Applied to electricity generation, $N$ describes the amount of electricity generated by a technology and $K$ describes total electricity consumption. The ratio $N/K$ therefore describes the relative share of a technology in the electricity supply.

There is an analytical solution for this equation – but we do not need it here. We can first identify the form of growth directly from the equation. If a technology is not yet very widespread, i.e. $N$ is small ($N \approx 0$), then the change is proportional to $r \, N$. This leads to exponential growth. If $N$ becomes increasingly widespread and approaches capacity, i.e., $N$ approaches $K$ ($N \approx K$), then the second term on the right-hand side approaches zero – that is, growth disappears and the innovation has become established.

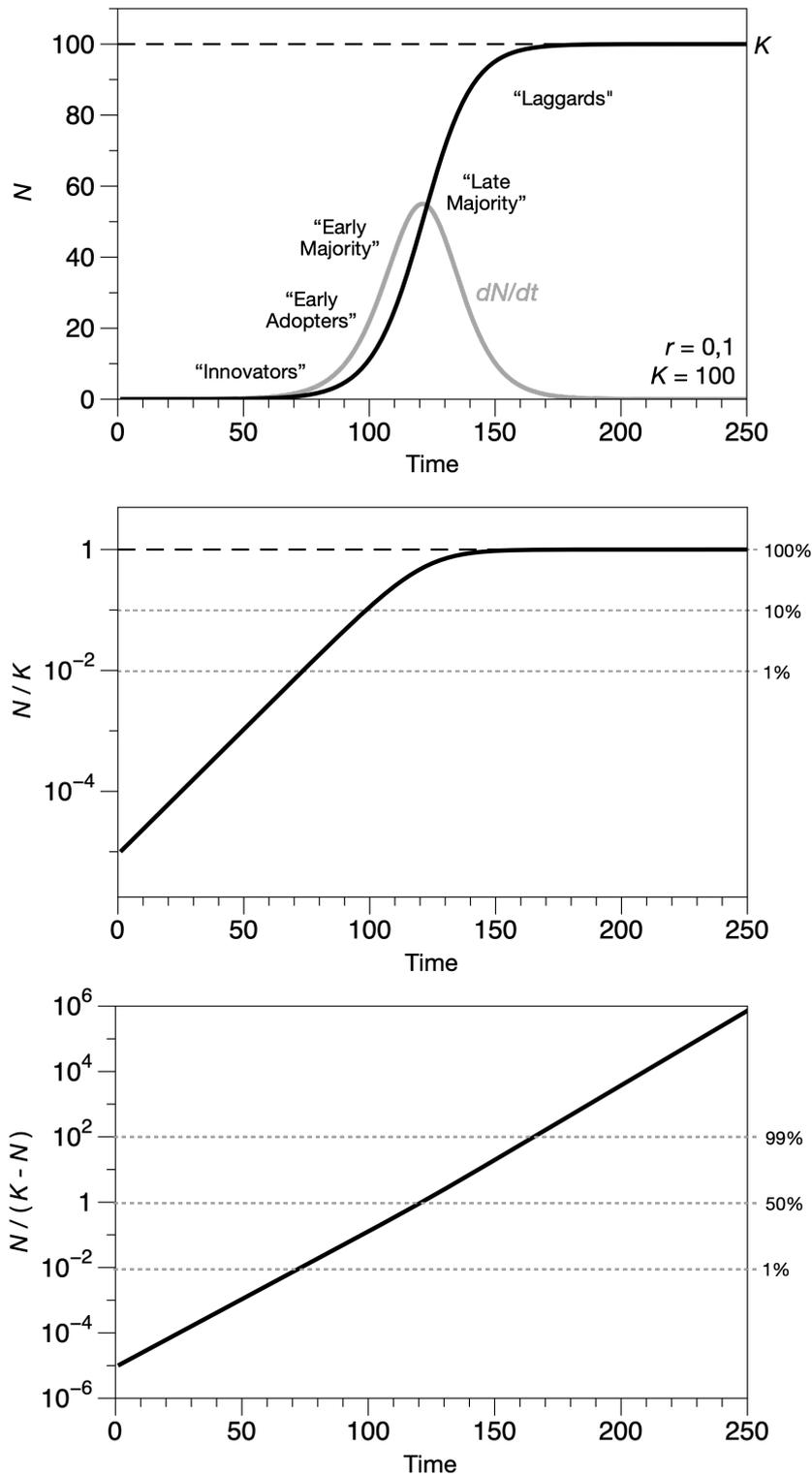

*Fig. 1 LOGISTIC GROWTH*
*Top) General representation of logistic growth of a variable N over time with the classification of growth areas in the diffusion of innovations using common terms. To represent logistic growth in data, it is helpful to not look at the ratio N/K (middle), but rather at the ratio N compared to the rest, K – N (bottom), because this ratio shows a consistently exponential function. The example shows the logistic curve with parameter values of r = 0.1 and K = 100.*

Why is such a description helpful when analysing data? It is because we as humans are notoriously bad at grasping exponential growth – in other words, we underestimate exponential growth. Logistic growth is initially exponential, although saturation ultimately occurs, meaning that $N$ approaches $K$. If we now look at only the share of a technology in the total, i.e., $N/K$, and plot this on a logarithmic axis, we see a kink in the curve. This occurs when a technology is already widely used and has almost saturated the market (Figure 1b). However, if we look at the share in relation to the rest, i.e., $N/(K - N)$, this leads to a purely exponential ratio (Figure 1c), which can be identified more clearly and easily in the data in the form of a linear relationship.

This simple transformation from the share $N/K$ to the modified share, $N/(K - N)$, and the representation with a logarithmic axis allows us to return the relationships to the familiar space of linear relationships. Let us now take a closer look at this relationship in the power generation data.

## Comparison of power generation trends

We use the EMBER dataset as our data source, which has compiled global data on electricity generation, installed capacity, and $CO_2$ emissions into a consistent dataset covering the years 2000 to 2024 [1]. We summarise the various technologies in four classes: first, fossil fuels (coal, oil, and gas); second, nuclear energy; third, photovoltaics and wind; and fourth, other renewable energy generation (especially hydropower and biomass). We then look at total electricity generation for different regions (in Terawatt hours per year, Figure 2, left side) and analyse it as logistic growth (Figure 2, right side), i.e., an analysis in which we set the share of each of the four classes in relation to the rest.

For Germany, total electricity generation has remained relatively stable over the last 25 years at around 600 TWh/a, although a slight decline has been evident since the end of the 2010s. We can see the phase-out of nuclear power in the decline in the red share in Figure 2, and fossil fuel-based forms of electricity generation (black) have also declined significantly. The share of photovoltaics (PV) and wind (blue), on the other hand, has grown over the years. If we now look at these trends in terms of logistic growth (Figure 2, right), the contribution of PV and wind is clearly recognisable as logistic growth. The share of PV and wind has therefore grown exponentially in relation to the rest of electricity generation over the last 25 years. In 2024, electricity generation from PV and wind exceeded the share of fossil fuel-based electricity generation for the first time, thus representing the dominant share of the four categories.

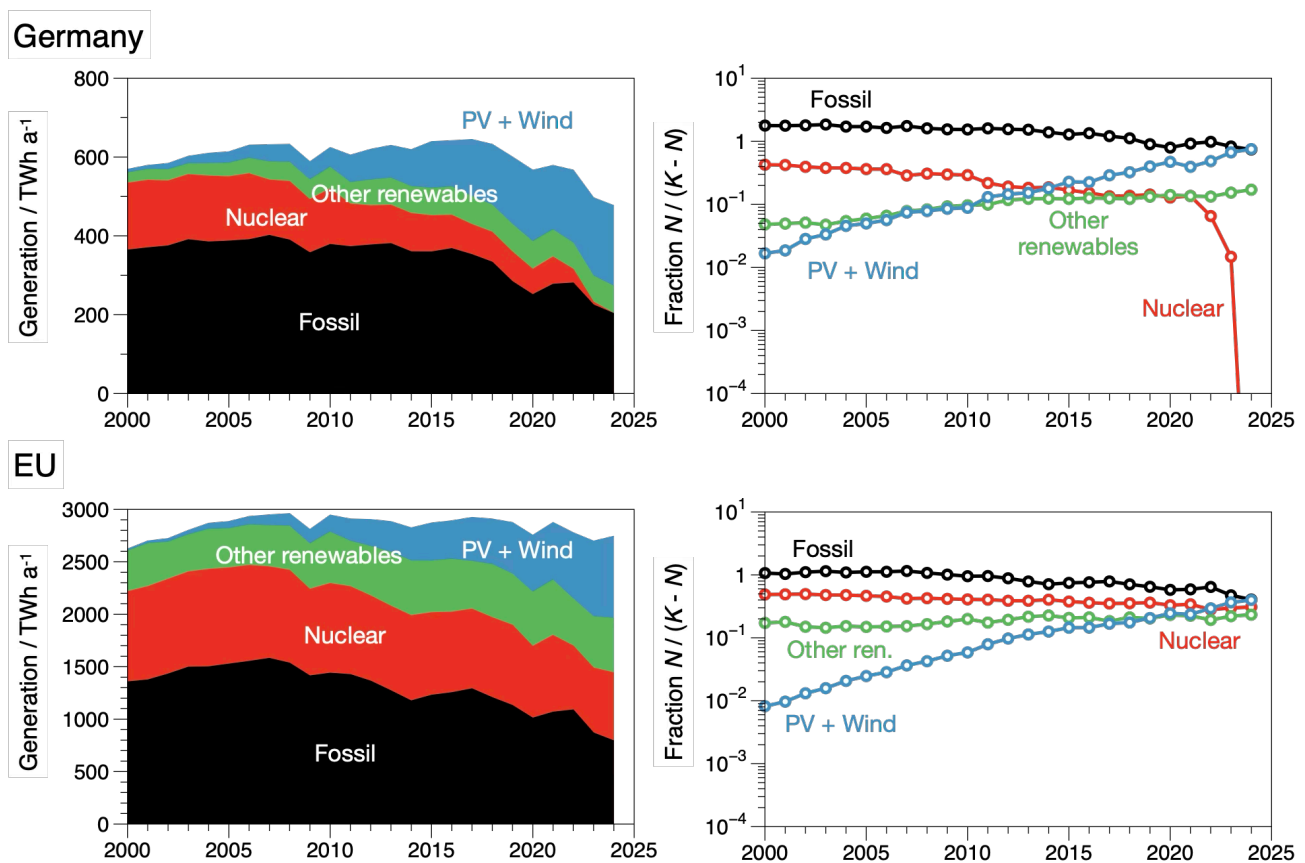

*Fig. 2 TRENDS IN ELECTRICITY GENERATION*
*Total annual electricity generation and the contributions of fossil fuel technologies (black), nuclear energy (red), other renewable energies (green), and photovoltaics and wind (blue) for a) Germany, b) Europe, c) the USA (next page), d) China (next page), e) India (next page), and f) worldwide (next page). The trends in the logistic growth of the four technologies are shown on the right-hand side.*

## USA

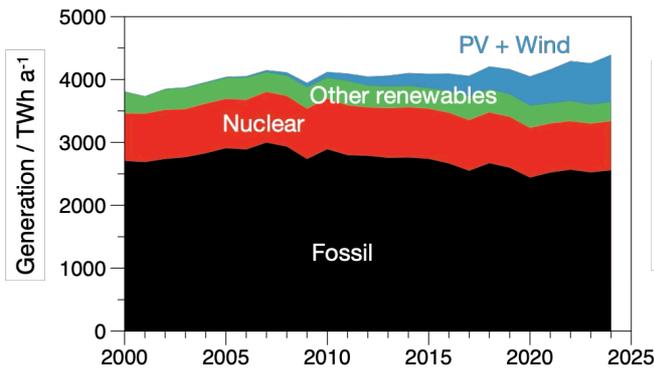
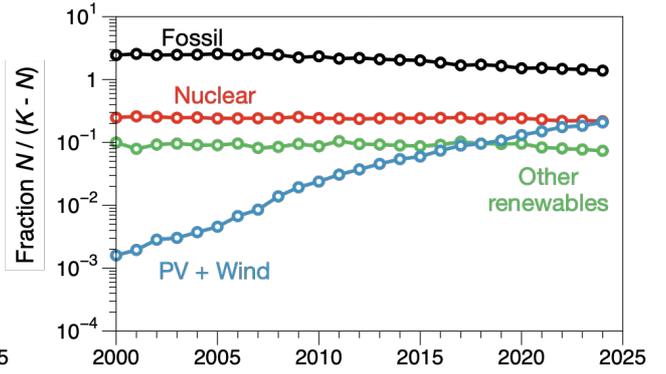

## China

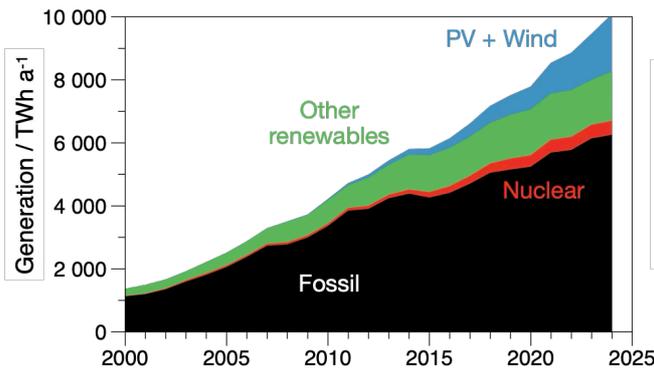
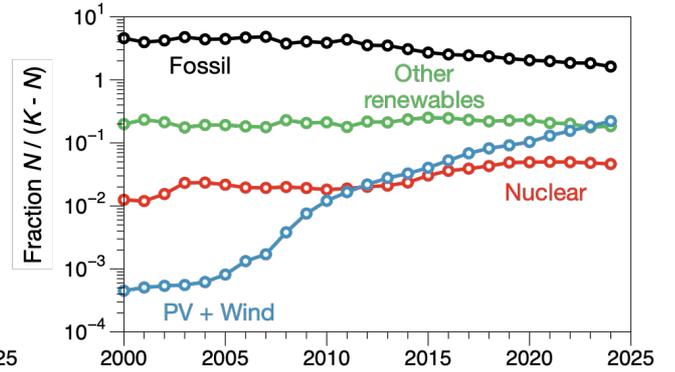

## India

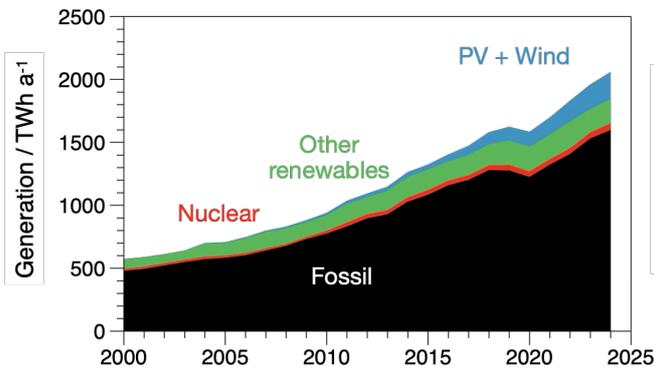
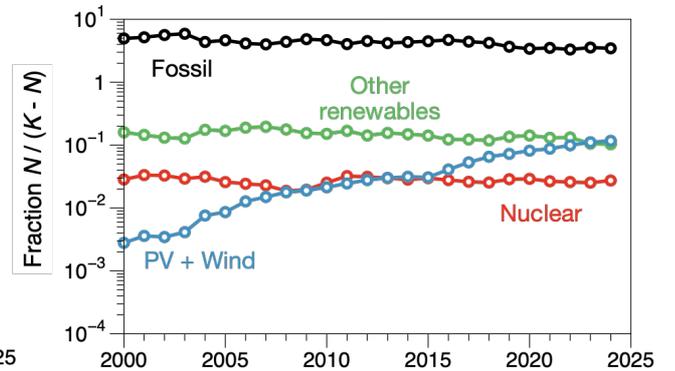

## Global

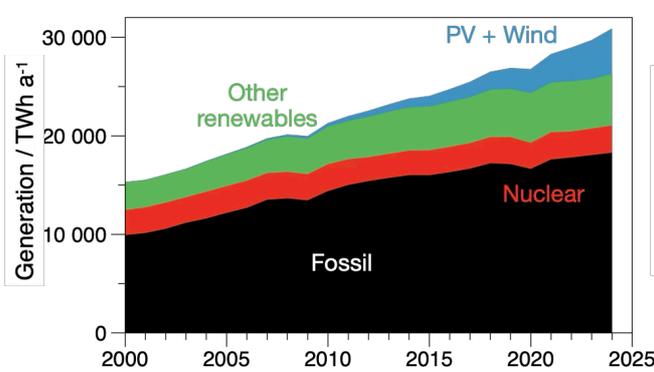
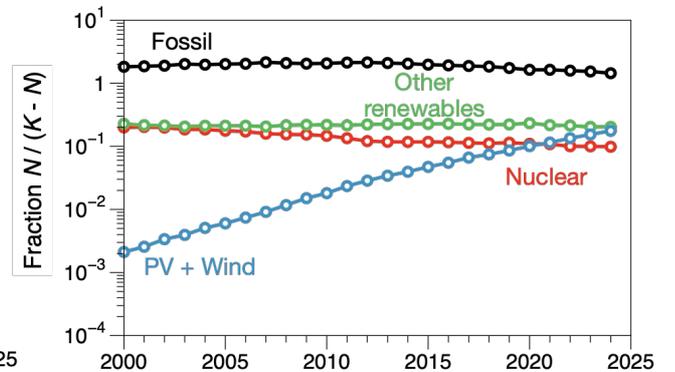

*Fig. 2 TRENDS IN ELECTRICITY GENERATION - continued*

Electricity generation across the EU and the US shows the same qualitative development. Electricity generation has remained relatively constant over the last 25 years at 2500-3000 TWh/a and 4000 TWh/a, respectively. The share of nuclear energy is generally higher, but has tended to fall slightly over the years, as has that of fossil fuel-based electricity generation. Once again, we see that PV and wind follow logistical growth, both in Europe and in the US. In the EU, PV and wind recently generated more electricity than nuclear energy and have roughly caught up with generation from fossil technologies. In the US, the share of fossil fuel-based electricity generation is significantly higher, and the share of PV and wind in electricity generation is not yet as high as in the EU or Germany. However, we see the logistic growth of PV and wind in all three cases.

The situation is somewhat different in India and China, which are still developing countries. This can also be seen in electricity generation: in India, it has risen from 600 TWh/a in 2000 to over 2000 TWh/a in 2024, and in China from just over 1000 TWh/a to over 10,000 TWh/a. As a result, the four technology classes have developed differently, with all of them growing over the years. In the description of logistic growth, this is reflected in a capacity $K$ that increases over time. When analysing the relative share in logistic growth, this increase in $K$ is eliminated by normalisation. As a result, we see that PV and wind also show logistic growth here. In relative terms, this shows that PV and wind are growing faster than other technologies in these countries, which is why their share is declining in percentage terms. This means, for example, that although nuclear energy and fossil fuel power generation continue to increase in China, power generation from PV and wind is growing significantly faster. In both countries, feed-in from PV and wind already surpassed power generation from nuclear energy more than ten years ago.

These trends are also evident in global developments (Figure 2): PV and wind grow logistically and thus displace other power generation technologies – even if this is not quite so obvious in terms of rising power generation.

**When do technologies become established?**

But why do renewable energies PV and wind grow significantly faster than other technologies, and why is this happening worldwide? Ultimately, this has to do with the cost of electricity generation. Photovoltaics and wind have now become the most cost-effective technologies for electricity generation [3]. This was not always the case. In 2000, PV and wind were still significantly more expensive. Since then, photovoltaics in particular has seen a positive price development, with costs falling significantly.

This dynamic can be described using a simple feedback diagram (Figure 3 left). At the beginning, an innovation is expensive—there is not yet sufficient demand. The reason: the costs are still too high. If the costs fall, this can increase demand. Higher demand leads to higher production—and higher production can, in certain circumstances, lead to lower manufacturing costs. In this case, the result is positive feedback: lower costs lead to higher demand. If demand is described mathematically as a function of cost, the derivative is negative, which is why there is a minus sign on the arrow in Figure 3 left. Similarly, higher demand leads to lower unit costs – again, the derivative of this mathematical relationship is negative. Overall positive feedback results when the product of the derivatives is positive, which is the case in our example. The strength of the feedback results from the product of the derivatives. With sharply lower unit costs at higher production rates, we therefore expect a stronger feedback.

This is precisely the dynamic that photovoltaics has shown. If we consider unit costs as a function of total production (Figure 3 right, from [4]), the significant reduction in manufacturing costs for photovoltaics – also referred to as a steep learning curve – is clearly evident. No such effect is found for the costs of nuclear energy in France. On the contrary, costs there have actually risen as more nuclear power plants have been built, which can be explained, among other things, by increased safety requirements. The enormous reduction in the cost of photovoltaics can be explained by the fact that it is a much simpler technology, where steep learning curves are more likely to be expected [5]. In other words, the breakthrough of PV and wind as global innovations in electricity generation is consistent with these dynamics. Due to these lower costs – and the feedback effect of learning curves, which lead to even lower costs – PV and wind are growing significantly faster and will dominate electricity generation in the long term.

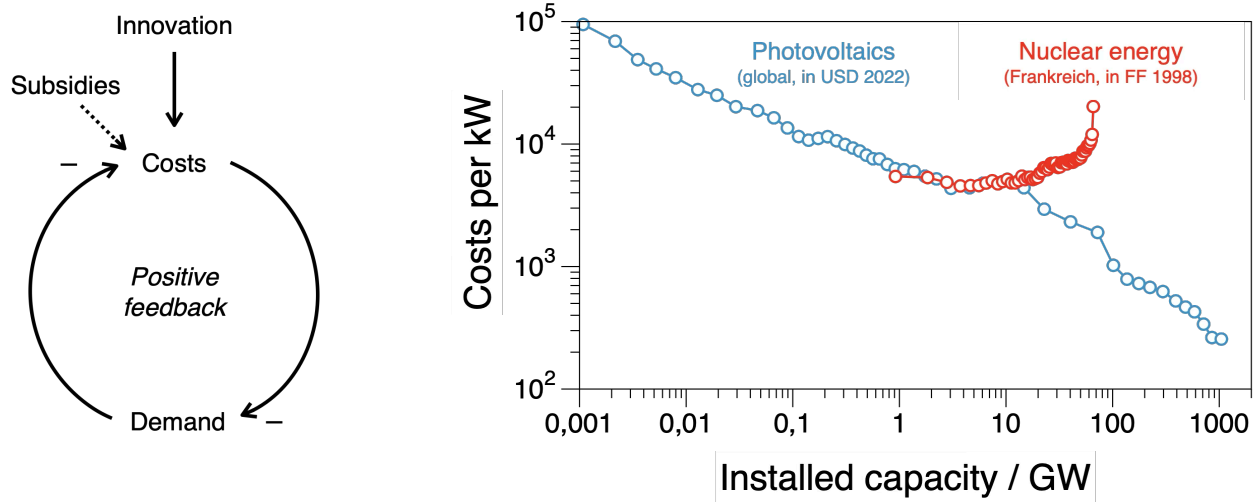

*Fig. 3 FEEDBACK AND LEARNING CURVES*
*(left) Schematic feedback diagram showing how initial subsidies can reduce the cost of an innovation, which can lead to higher demand, which can reduce unit costs in production. Whether such feedback occurs depends on the learning curve of a technology, i.e., whether more demand leads to lower unit costs. (right) This is observed in photovoltaics, but not in nuclear energy (learning curves from [4]).*

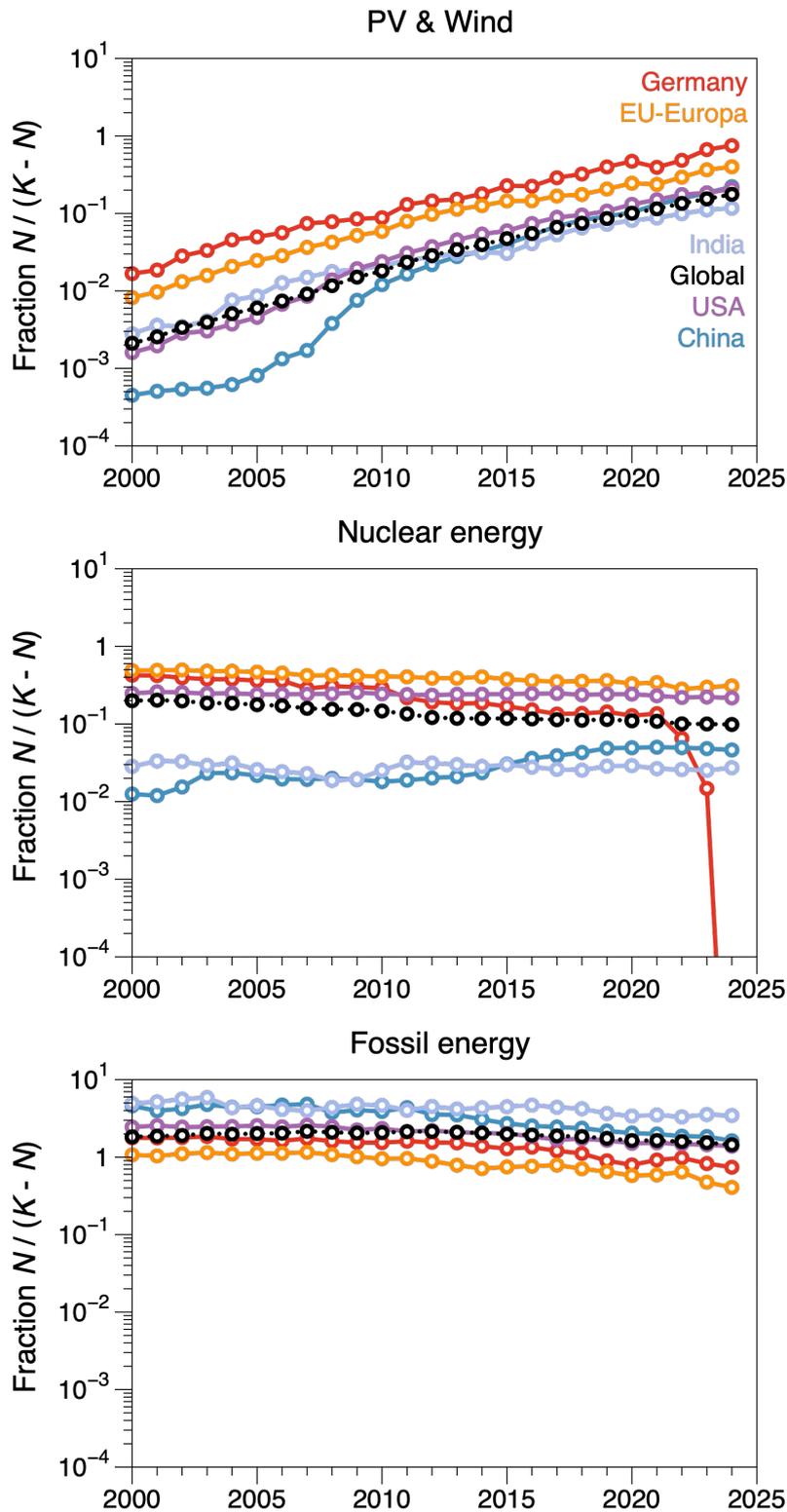

*Fig. 4 COMPARISON OF ELECTRICITY TRENDS*
*The development over time of the shares of (top) photovoltaics and wind, (middle) nuclear energy, and (bottom) fossil fuels compared to the rest of electricity generation for the selected regions.*

## Technology change in comparison

In summary, we see that photovoltaics and wind energy will dominate global electricity generation—simply because they are the cheapest and have steep learning curves. It is astonishing how parallel the developments are (Figure 4 top). This probably reflects the fact that everything is included in this resulting "mishmash" of technology diffusion: technology functionality and costs, consumer and worker education, regulatory developments, capital investments, etc. This pattern suggests that while different environments, regions, and legal systems may flavour the dish with local spices, as is the case with climate, we are all eating the same "energy dish." The most important and probably most difficult decision that each national chef had to make was when to start.

This comparison also shows that Germany is playing a pioneering role—in our electricity generation, PV and wind have already reached the highest share in the regions we have compared here. Germany is not taking a wrong turn or a special path. Rather, these trends—logistical growth of PV and wind, as well as the decline in the relative contribution of fossil fuels and nuclear energy—are quite similar worldwide (Figures 4 middle, bottom). Only in the case of nuclear energy is it noticeable that China is still expanding it somewhat, with its share rising slightly (blue line in Figure 4 middle), but not nearly to the extent that PV and wind grow.

In this context, we can now also classify the problems in the German electricity market. Subsidies help to initiate this development (Figure 3 left). These can initially reduce costs, allowing demand to rise and unit costs to fall. This enables positive feedback. This is exactly what happened in the early 2000s with the Renewable Energy Sources Act ("EEG"). At that time, PV and wind were still significantly more expensive. They were promoted through subsidies, which reduced the electricity production costs of PV and wind worldwide.

However, we are still paying part of the subsidies from that time today—these are the costs that the pioneers of innovation have to bear, but in return they have the advantage of developing solutions when the technology reaches market saturation. The latter is reflected in the fact that, in percentage terms, the feed-in from fluctuating renewables PV and wind is highest in our country. This makes it understandable that grid expansion and the expansion of electricity storage facilities are most urgent. This enormous opportunity, which has to do with the technological development of such a novel electricity system, can be exploited through good, forward-looking energy policy.

## Summary

*Photovoltaics and wind energy are on the rise worldwide. Electricity generation data from various regions over the last 25 years show that this rise follows a logistical growth curve that progresses at an astonishingly uniform rate. Germany does not play a special role in this, but is among the pioneers of this development. Fossil fuel power generation and nuclear power are being displaced as power generation technologies. This can be explained by the fact that photovoltaics and wind are comparatively simple technologies with steep learning curves and already have the lowest electricity production costs.*

## Keywords

## The authors

*Axel Kleidon studied physics and meteorology at the University of Hamburg and Purdue University, Indiana, USA. After completing his doctorate at the Max Planck Institute for Meteorology, he conducted research at Stanford University in California and the University of Maryland. Since 2006, he has headed the "Theory and Modeling of the Biosphere" group at the Max Planck Institute for Biogeochemistry in Jena. His research interests range from the thermodynamics of the Earth system to the natural limits of renewable energy sources.*

*Harald Lesch studied physics and philosophy at the Universities of Giessen and Bonn and received his doctorate from the University of Bonn. He then conducted research in Heidelberg and Toronto. After completing his habilitation at the University of Bonn in 1994, he became a full professor of theoretical astrophysics at LMU in 1995 and, since 2002, has been a lecturer in natural philosophy at the Munich School of Philosophy. His work covers complex cosmic and terrestrial systems, including the natural limits of technological societies.*

*Russ Conser studied mechanical engineering at Iowa State University and spent thirty years at Shell, where he started out in global oil and gas production, then moved into corporate strategy, and finally headed up investment, innovation, and research in the field of energy technology. Before retiring in 2013, he led Shell's global innovation program, GameChanger, which gave him early insight into the emerging energy landscape. He is currently co-founder and CEO of Standard Soil and a venture partner at Ecosphere Ventures, a company that invests in transformative climate technologies and sustainability solutions.*

***Address***
*Dr. Axel Kleidon, Max Planck Institute for Biogeochemistry, P.O. Box 10 01 64, 07701 Jena, Germany.* akleidon@bgc-jena.mpg.de

**Internet**
*Who will dominate the energy future? A video accompanying this article on YouTube*

https://youtu.be/aJXTH65VyAk